\documentclass{emulateapj}

\usepackage{natbib}
\begin{document}

\newcommand{\qsoa}{J0256+0019}
\newcommand{\qsoafull}{SDSS~J025617.7+001904}
\newcommand{\qsob}{J0050+3445}
\newcommand{\qsobfull}{CFHQS~J005006.6+344522}
\newcommand{\br}{BR~1202-0725}

\newcommand{\Msunyr}{M_{\sun}~{\rm yr}^{-1}}
\newcommand{\fcgs}{{\rm erg}~{\rm s}^{-1}~{\rm cm}^{-2}~{\rm \AA}^{-1}}
\newcommand{\ergss}{{\rm erg}~{\rm s}^{-1}}
\newcommand{\kms}{{\rm km}~{\rm s}^{-1}}
\newcommand{\Lstar}{L^{*}}

\newcommand{\galfit}{GALFIT}
\newcommand{\astrodrizzle}{\textsc{AstroDrizzle}}
\newcommand{\sextractor}{SExtractor}

\shorttitle{QSO companions}
\shortauthors{McGreer et al.}

\title{Close companions to two high-redshift quasars\altaffilmark{\dag}}

\altaffiltext{\dag}{Based in part on observations made with the NASA/ESA Hubble Space Telescope, obtained at the Space Telescope Science Institute, which is operated by the Association of Universities for Research in Astronomy, Inc., under NASA contract NAS 5-26555. These observations are associated with programs \#12184 and \#12493. Observations were also made with the LBT and MMT.}

\author{
Ian D. McGreer,\altaffilmark{1}
Xiaohui Fan,\altaffilmark{1}
Michael A. Strauss,\altaffilmark{2}
Zolt\`an Haiman,\altaffilmark{3}
Gordon T. Richards,\altaffilmark{4}
Linhua Jiang,\altaffilmark{5, \ddag}
Fuyan Bian,\altaffilmark{1,6} and
Donald P. Schneider\altaffilmark{7}
}
\altaffiltext{1}{Steward Observatory, The University of Arizona, 
                 933 North Cherry Avenue, Tucson, AZ 85721--0065}
\email{imcgreer@as.arizona.edu}
\altaffiltext{2}{Princeton University Observatory, Peyton Hall,
                 Princeton, NJ 08544, USA.}  
\altaffiltext{3}{Department of Astronomy, Columbia University, 
                 550 West 120th Street, New York, NY 10027, USA.}
\altaffiltext{4}{Department of Physics, Drexel University, 
                 3141 Chestnut Street, Philadelphia, PA 19104, USA.}
\altaffiltext{5}{School of Earth and Space Exploration, 
                 Arizona State University, Tempe, AZ 85287, USA.}
\altaffiltext{6}{Research School of Astronomy and Astrophysics, 
                 Australian National University, 
                 Canberra, ACT 2611, Australia.}
\altaffiltext{7}{Department of Astronomy and Astrophysics and the Institute for Gravitation and the Cosmos, 
                 The Pennsylvania State University,
                 University Park, PA 16802, USA.} 
\altaffiltext{\ddag}{Hubble Fellow}
                 
\begin{abstract}
We report the serendipitous discoveries of companion galaxies to two 
high-redshift quasars. \qsoafull\ is a $z=4.79$ quasar included in our recent 
survey of faint quasars in the SDSS Stripe 82 region. The initial MMT slit
spectroscopy shows excess Ly$\alpha$ emission extending well beyond the
quasar's light profile. Further imaging and spectroscopy with LBT/MODS1 
confirms the presence of a bright galaxy ($i_{\rm AB} = 23.6$) located 
2\arcsec\ (12~kpc~projected) from the quasar with strong Ly$\alpha$ emission 
(EW$_0 \approx 100$~\AA) at the redshift of the quasar, as well as faint 
continuum. The second quasar, \qsobfull\ ($z=6.25$), is included in our recent 
HST SNAP survey of $z\sim6$ quasars searching for evidence of gravitational
lensing. Deep imaging with ACS and WFC3 confirms an optical dropout 
$\sim4.5$~mag fainter than the quasar ($Y_{\rm AB}=25$) at a separation of 
0.9\arcsec. The red $i_{775}-Y_{105}$ color of the galaxy and its proximity to 
the quasar (5~kpc~projected if at the quasar redshift) strongly favor an 
association with the quasar. Although it is much fainter than the quasar it is 
remarkably bright when compared to field galaxies at this redshift, while 
showing no evidence for lensing. Both systems may represent late-stage 
mergers of two massive galaxies, with the observed light for one dominated by 
powerful ongoing star formation and for the other by rapid black hole growth. 
Observations of close companions are rare; if major mergers are primarily 
responsible for high-redshift quasar fueling then the phase when progenitor 
galaxies can be observed as bright companions is relatively short.
\end{abstract}

\section{Introduction}

The discovery of quasars as distant as $z\sim7$ already powered by 
black holes (BHs) with masses $\sim10^9~M_{\sun}$ 
\citep{Fan+01PI,Mortlock+11,Venemans+13} presents a challenge for early 
structure formation models. First, some physical process is needed to 
generate seed BHs at even higher redshifts 
\citep[e.g.,][]{Volonteri10,Haiman13}. 
Next, regardless of the nature of the seeds, the initial BHs must grow 
rapidly in order to reach a billion solar masses or more in less than a Gyr.
$\Lambda$CDM models generally predict that this high redshift growth
occurs by gas accretion rather than by a succession of black 
hole mergers \citep{Li+07,DiMatteo+08,TH09}, 
requiring both a enormous fuel supply and
a mechanism for driving the gas from intergalactic scales down to the
central regions of galaxies where it can be accreted onto the BH.

Mergers of gas-rich disk galaxies have been proposed as a solution
to this problem \citep[e.g.,][]{KH00,Hopkins+06,Li+07,Hopkins+08}. 
In the merger scenario the remnant typically passes through an obscured 
starburst phase followed by a luminous, unobscured quasar phase. Merger 
dynamics provide a natural mechanism for shedding angular momentum and 
driving gas to central regions \citep{Hernquist89}, plausibly accounting 
for the continuous fuel supply required to grow high-redshift quasars.

While the merger hypothesis is consistent with many observations, direct 
evidence for merger activity associated with individual quasars is rare.
This is  likely because the observational signatures of a recent merger 
\citep[tidal tails, gas shells, etc.; see, e.g.,][]{Bennert+08} are 
short-lived and faint, thus easily overwhelmed by the luminous quasar. 
Nonetheless, early Hubble Space Telescope (HST) observations showed that at 
least some nearby luminous quasars show evidence for recent interactions, 
with a significant fraction  having close ($<10$~kpc) companion galaxies 
\citep{Bahcall+97}. In addition, the fraction of highly dust-obscured quasars 
showing evidence of recent merger activity is close to unity  
\citep{ULB08}, in agreement with the general outline of the merger scenario.

\begin{figure*}
 \epsscale{1.15}
 \plotone{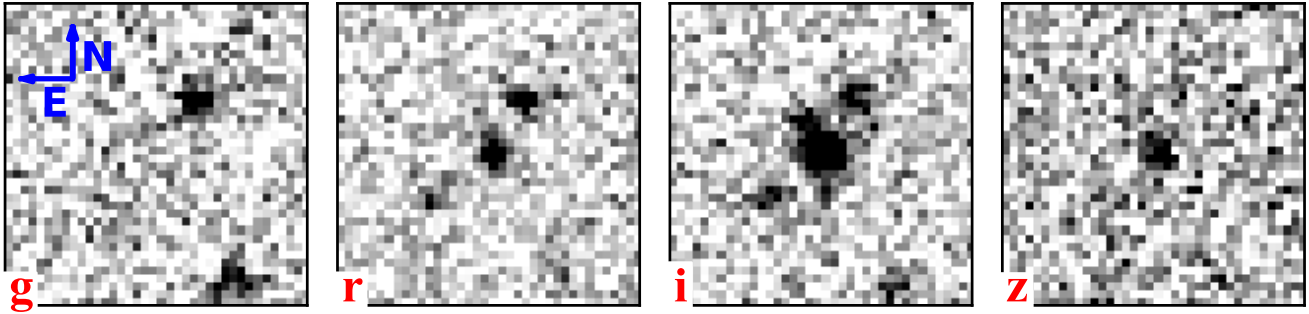}
 \caption{SDSS deep Stripe 82 $griz$ images of \qsoa. Images are 20\arcsec\ 
 on a side with a pixel scale of 0\farcs{45} and are oriented with North up 
 and East to the left. The companion  galaxy is visible 2\arcsec\ NE of the 
 quasar in the $i$-band image.
 Two additional galaxies are detected to the NW and SE at distances of
 $\sim5$\arcsec\ from the quasar and are likely at low redshift given their 
 relatively blue colors.
 \label{fig:J0256griz}
 }
\end{figure*}

Mergers are more likely to occur in overdense environments where the rate of 
galactic encounters is enhanced. Even if direct evidence for recent merging 
activity in quasar host galaxies is elusive, circumstantial evidence in the 
form of local overdensities may argue in favor of merger-driven growth. 
For example, in the merger-tree-based hydrodynamical simulation of 
\citet{Li+07}, a single $z\sim6.5$ quasar is assembled from a succession of 
seven major mergers, so that the quasar is surrounded by nearby ($\la20~$kpc) 
companions more or less continuously from $z\sim12$ to $z\sim7$ 
\citep[see Fig.~6 of][]{Li+07}. 
More broadly, the expectation that exceedingly rare high-redshift quasars 
would occupy the most highly biased regions at their 
observed epoch has motivated many surveys of their environments on few Mpc 
scales \citep{Stiavelli+05,Willott+05,Zheng+06,Kashikawa+07,Kim+09,Utsumi+10,
Benitez+13,Husband+13}, the results of which have been inconclusive, with 
some quasar fields having apparent overdensities of candidate associated 
galaxies, and some not. Some recent theoretical models even suggest that 
high-redshift quasars do {\em not} inhabit the strongest large-scale 
overdensities at high redshift \citep{Fanidakis+13}.

A particularly well-studied case of prodigious merger activity associated
with a luminous, high-redshift quasar is \br\ at $z=4.7$. One of the first 
detections of a FIR-hyperluminous, high-redshift sub-mm galaxy (SMG)
resulted from observations of the host galaxy of this quasar \citep{Isaak+94}.
Subsequently, an optically undetected SMG as bright as the quasar host at 
sub-mm wavelengths was discovered a mere 4\arcsec\ from the quasar 
\citep{Omont+96}, as was a pair of Ly$\alpha$-emitting galaxies within 
$\sim2$\arcsec\ \citep{HME96,Petitjean+96}. 
\br\ has been dubbed the ``archetypal'' 
system of a close group of galaxies leading to mergers and fueling both SMBH 
formation and prodigious star formation \citep{Carilli+13}. 
It is curious, however, that 
observations of systems like \br\ are rare, if indeed group-scale mergers
are a key pathway to forming high-redshift quasars.

In this paper we report the discovery of close companions ($\la10$~kpc)
of two high-redshift quasars. The first example is a companion galaxy to the
$z\sim5$ quasar \qsoafull~(hereafter \qsoa). This galaxy is located 2\arcsec\ 
from the quasar and has strong Ly$\alpha$ emission at the same redshift. It 
bears many similarities to the Ly$\alpha$-emitting companions of \br. We
present imaging and spectroscopy of this system in \S\ref{sec:j0256}.
The second quasar, \qsobfull~(hereafter \qsob), was included in an HST SNAP 
survey of $z\sim6$ quasars searching for evidence of gravitational lensing.
A galaxy 0\farcs{9} from the quasar was detected in the HST image.
In \S\ref{sec:j0050} we present strong evidence from HST imaging that this 
galaxy is almost certainly associated with the quasar. In \S~\ref{sec:discuss} 
we consider the physical origin of the observed emission from both galaxies, 
discuss these observations in the context of the merger hypothesis, and draw 
rough conclusions on the incidence of close companions for similar quasars. 
We present brief conclusions and speculate on the nature of companion galaxies
in \S\ref{sec:conclude}.

All magnitudes are on the AB system \citep{OkeGunn} and have been 
corrected for Galactic extinction using the \citet{SFD98} extinction 
maps unless otherwise noted. We adopt a $\Lambda$CDM cosmology with 
parameters $\Omega_\Lambda=0.727$,~$\Omega_m=0.273$,~
$\Omega_{\rm b}=0.0456$,~and~$H_0=70~{\rm km}~{\rm s}^{-1}~{\rm Mpc}^{-1}$
\citep{Komatsu+11} when needed.

\begin{figure}
 \epsscale{1.15}
 \plotone{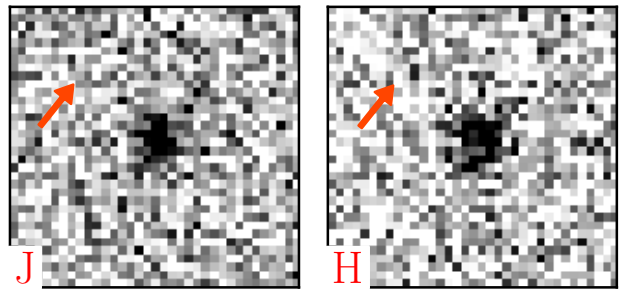}
 \caption{MMT SWIRC $J$ and $H$ images of \qsoa. Images are 5\arcsec\ on
 a side with a pixel scale of 0\farcs{15} and are oriented with North up 
 and East to the left. The position of the companion galaxy is marked
 with an arrow.
 The quasar is detected in both images but the companion galaxy is not.
 \label{fig:J0256swirc}
 }
\end{figure}

\section{\qsoa}\label{sec:j0256}

The quasar \qsoa\ was selected as a $z\sim5$ quasar candidate based on optical 
and near-IR color selection \citep{McGreer+13}. The input imaging included 
deep coadded $ugriz$ \citep{Fukugita+96} images from the Sloan Digital Sky 
Survey \citep[SDSS][]{York+00} Stripe 82 region, with $\sim80\mbox{--}100$ 
individual images contributing to the coadd at each position 
(see \citealt{Jiang+09}~and~\citealt{Jiang+14} for details).
These images reach $\sim2$~mag deeper than single-epoch SDSS imaging.
$J$-band imaging from the UKIRT Infrared Deep Sky Survey 
\citep[UKIDSS; ][]{Lawrence+07} was also used for the color selection.
In this section we describe subsequent observations of the quasar and
companion galaxy using MMT and LBT.

\subsection{MMT Observations}

\qsoa\ was confirmed as a $z=4.8$ quasar using longslit spectroscopic
observations obtained with the MMT Red Channel spectrograph on 2011 Oct 01. 
The Red Channel observations were performed with a 
1\arcsec$\times$180\arcsec\ slit and the low dispersion 
270~mm$^{-1}$ grating, delivering a resolution of $R \sim 640$ over the 
range 5700~\AA\ to 9700~\AA. Three 10m exposures were combined
to produce the final spectrum. Data were processed in a standard fashion; 
details are given in \citet{McGreer+13}.

\begin{figure}
 \epsscale{1.15}
 \plotone{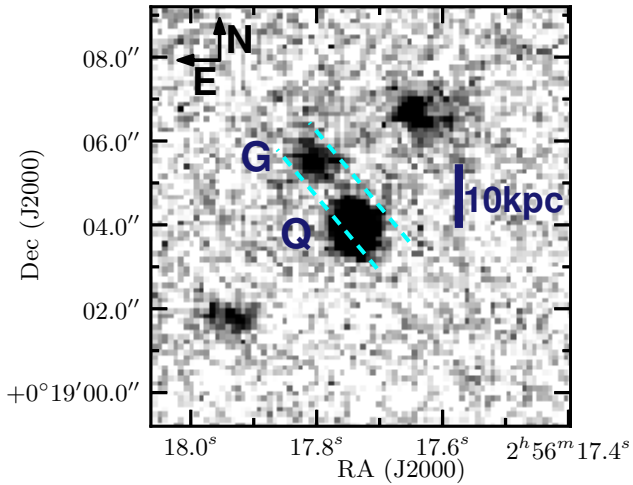}
 \caption{LBT/MODS $i$-band image of \qsoa. 
 The scale is 0.123\arcsec\ per pixel and the seeing is $\sim0.6$\arcsec. 
 The companion galaxy (labeled G) is easily resolved from the quasar (Q) and 
 is extended compared to stars within the field. The galaxy has $i=23.6$.
 The dashed cyan lines correspond to the orientation of the slit during the
 spectroscopic observations.
 \label{fig:J0256LBTi}
 }
\end{figure}

We immediately noticed extended Ly$\alpha$ emission in the two-dimensional
spectrum. Although the seeing during the spectroscopic observations was 
$\sim0.8$\arcsec, the emission in the vicinity of the quasar's Ly$\alpha$ 
line extends to roughly 2\arcsec.
The Ly$\alpha$ detection was purely serendipitous: only after the 
observations did we notice extended emission in the same location in the 
stacked $i$-band image from Stripe 82 (Figure~\ref{fig:J0256griz}). 
Although we applied no morphological criteria in selecting quasar candidates 
from Stripe 82, \qsoa\ would have likely been targeted anyway, as the 
$i$-band detection for the quasar has \sextractor\ CLASS\_STAR=0.95 (where 1 
represents a stellar profile); the excess $i$-band emission 
is quite weak relative to the quasar.

We obtained near-IR imaging in the $J$ and $H$ bands with the
MMT Smithsonian Widefield Infrared Camera \citep[SWIRC;][]{swirc}
on 2011 Oct 15. The seeing was 0.7\arcsec\ and conditions were variable
and non-photometric. \qsoa\ was observed in the $J$ and $H$ bands using
dithered integrations for a total exposure time of 20.7m and 16.3m, 
respectively.
The images were shifted and stacked,
and then registered to an astrometric solution obtained by matching UKIDSS
stars within the field.
Flux calibration was obtained from observations of faint UKIRT standard
stars from \citet{Leggett+06} and checked against UKIDSS matches. 
The quasar is clearly detected in both images, with fluxes of 
$J = 22.11 \pm 0.14$ and $H = 21.43 \pm 0.10$ measured from 
1.8\arcsec\ apertures (Figure~\ref{fig:J0256swirc}).
The companion object is not detected to $3\sigma$ limits of $J=22.8$ 
and $H=23.2$ (both AB).

\subsection{LBT/MODS Observations}

Additional imaging and spectroscopic observations were obtained with the 
Large Binocular Telescope (LBT) Multi-Object Double Spectrograph (MODS1) 
instrument on 2012 Sep 21. MODS1 is an optical imager/spectrograph 
mounted on the LBT 2$\times$8.4m telescope \citep{MODS}.
We obtained both direct imaging and grating spectroscopy of \qsoa.
During the observations conditions were non-photometric with mostly clear
skies and good seeing, improving from 0.8\arcsec\ to 0.6\arcsec.

\subsubsection{Imaging}\label{sec:lbtimaging}

In acquisition mode the MODS1 red CCD employs a 1024$\times$1024 
array with a pixel scale of 0.123\arcsec\ and a FOV of 2.1\arcmin.
Immediately following a series of longslit spectroscopic observations 
(see below), a single 600s exposure was obtained with the SDSS $i$ filter. 
Guiding was continuous throughout both observations.

The $i$-band image was reduced in a standard fashion.
A bias correction was applied from a median of bias images. Sky flats 
were obtained at twilight and used to provide the flat field. 
The image was masked of regions affected by scattered light from a nearby
bright star and from obstruction by the guide probe camera, leaving
roughly 2/3 of the field usable.
Astrometric registration and flux calibration were obtained by matching 
stellar objects from the Stripe 82 coadded imaging.
Object detection was performed with \sextractor\ \citep{sextractor}
using standard parameters.

\begin{figure}
 \epsscale{1.15}
 \plotone{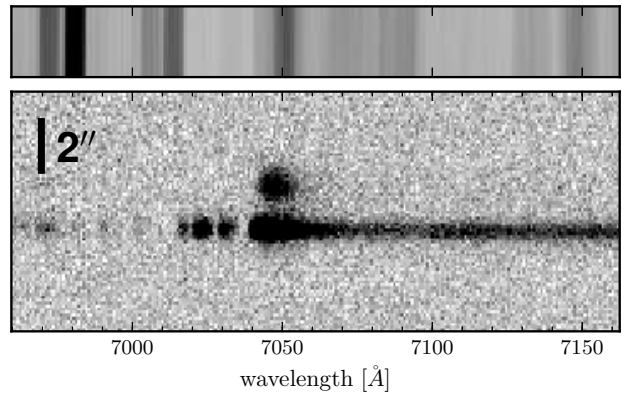}
 \caption{
  Two dimensional LBT/MODS spectrum showing the Ly$\alpha$ emission region 
  (seeing $\sim0.8$\arcsec). The image in the top panel is the model for the
  sky background, and in the bottom panel the sky-subtracted spectrum. 
  Transmission through the Ly$\alpha$ forest as well as the rest-frame UV 
  continuum redward of Ly$\alpha$ are visible for the quasar, while the 
  companion presents only strong Ly$\alpha$ emission.
 \label{fig:J0256LBT2D}
 }
\end{figure}

\begin{figure*}
 \epsscale{1.15}
 \plotone{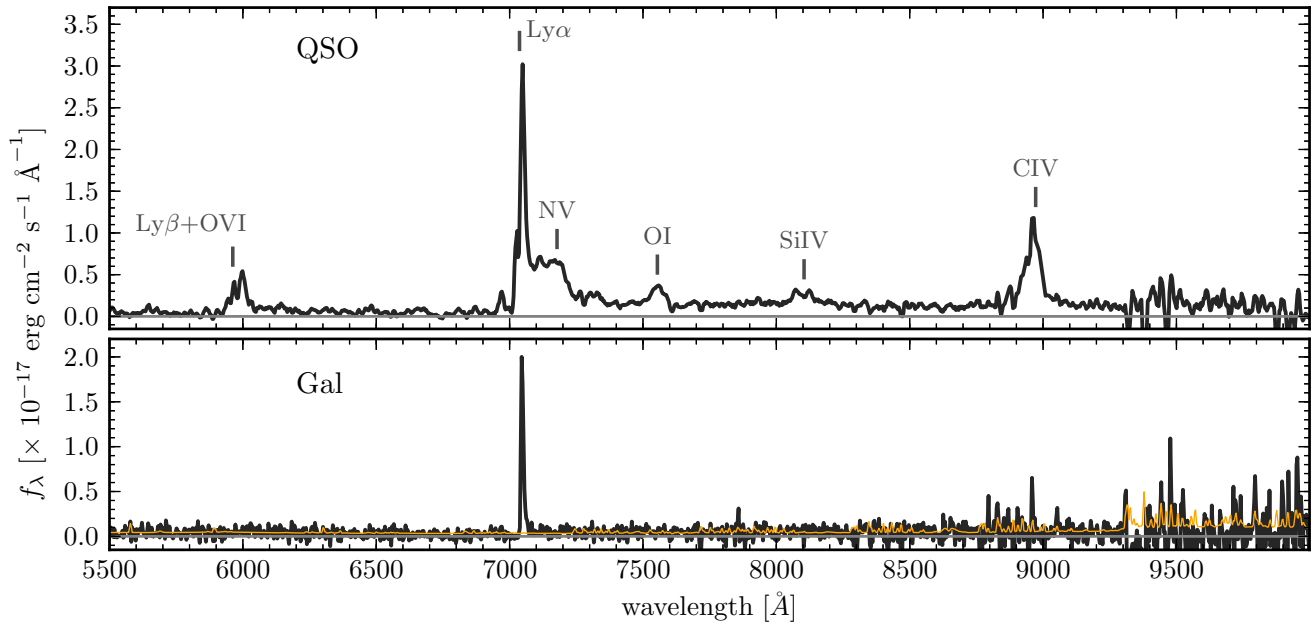}
 \caption{Extracted one dimensional LBT/MODS spectra. The spectra have been
 corrected for slit losses (\S\ref{sec:lbtspec}),
 rebinned to a dispersion of 3.4~\AA~pix$^{-1}$, and smoothed with a 
 4-pixel boxcar for display purposes. The thin orange line shows the 
 $1\sigma$ rms noise per pixel . The locations of typical quasar emission 
 lines (using a redshift derived from a Gaussian fit to the \ion{O}{1} line) 
 are marked in the quasar spectrum.
 \label{fig:J0256LBTspec}
 }
\end{figure*}

We measure a separation between the quasar and the companion object of
1.8\arcsec\ from the MODS1 $i$-band image, corresponding to a projected 
physical separation of 12.1~kpc for our adopted cosmology. The quasar has 
$i=22.08~\pm~0.02$, somewhat fainter than measured from the Stripe 82 coadded 
imaging ($i=21.73~\pm~0.02$). This is likely due to blending, although 
variability could also contribute to the flux difference. The two objects are
well resolved in the MODS1 image (Figure~\ref{fig:J0256LBTi}). According to the
\sextractor\ star/galaxy separation, the quasar is consistent with a point 
source (CLASS\_STAR = 0.97) while the companion is extended (CLASS\_STAR = 
0.17). The companion has $i=23.60~\pm~0.07$ and measured size of 
FWHM=1.0\arcsec\ (compared to 0.57\arcsec\ for the quasar). The companion is 
not detected in the Stripe 82 coadded $z$-band image to a $3\sigma$ limit of 
$z=23.5$; so $i-z \la 0.1$ for this object.

\subsubsection{Spectroscopy}\label{sec:lbtspec}

In grating spectroscopic mode the MODS1 Red channel utilizes a 8k$\times$3k
array. We used the single grating mode with the G670L grating and a 
1\arcsec\ longslit, providing a spectral resolution of $R\sim1400$ and 
wavelength coverage from 5800\AA\ to 1$\mu$m at a dispersion of 
0.8~\AA~pix$^{-1}$. Three exposures of 1200s each were obtained.

The position angle between the quasar and the companion is $31.2^\circ$ from 
the MODS1 imaging. We estimated a PA of $40^\circ$ from the Stripe 82 coadded
imaging and placed the MODS longslit at this angle for the spectroscopic
observations. 
Although the slit angle was offset by $\sim9^\circ$
from the true PA between the two objects, the quasar was also slightly 
off-center relative to the slit, while the companion galaxy was relatively
well-centered. 

The images were first processed with the MODS CCD reduction 
utilities (modsTools v0.3) to obtain bias and flat-field corrections.
Transformations between image pixels and wavelength were determined
from polynomial fits to arc calibration images.
A two-dimensional sky model was fit to each image using b-splines 
\citep{Kelson03} and then subtracted. Cosmic rays were identified
and masked during the construction of the sky model. The individual
exposures were combined with inverse variance weighting to produce
the final 2D spectrum.

Figure~\ref{fig:J0256LBT2D} displays a section of the sky-subtracted and 
combined two-dimensional spectrum obtained with MODS1. Two components 
are clearly visible at $\sim7045$\AA\ and are well separated. The lower
component is the quasar \qsoa\ included in our Stripe 82 survey. 
The only obvious feature in the galaxy
spectrum (upper component) is prominent Ly$\alpha$ emission.

\begin{figure}
 \epsscale{1.15}
 \plotone{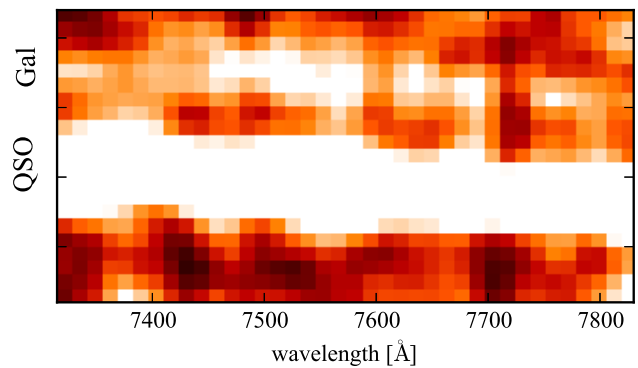}
 \caption{Two dimensional LBT MODS spectrum highlighting the faint continuum
 detected from the companion galaxy. The spectrum has been rebinned by a 
 factor of two along the spatial axis (0.25\arcsec~pix$^{-1}$) and a factor 
 of 16 along the spectral axis (14\AA~pix$^{-1}$).
 The image has been further smoothed by a Gaussian filter with $\sigma=1$ pix
 to enhance the faint continuum emission from the companion galaxy.
 \label{fig:J0256LBTcont2D}
 }
\end{figure}

One dimensional spectra of both components were obtained using optimal 
extraction \citep{Horne86} while allowing for multiple, blended object 
profiles \citep{Hynes02}.  We used the direct image obtained immediately 
following the spectroscopic observations to obtain the profile weights for 
extraction. We fit a Moffat profile to the quasar and an exponential profile
to the galaxy, then collapsed the profile fits along the slit axis.
The results are similar to those obtained with a simple boxcar extraction,
but with lower noise as expected.

Flux calibration is provided by observations of the spectrophotometric 
standard star GD71. This calibration is affected by the non-photometric 
conditions and slit losses. We obtain rough corrections for the slit
losses using the direct imaging obtained immediately after the spectroscopic 
observations, by calculating the ratio of the synthetic fluxes obtained from 
integrating the spectra to the total fluxes measured from the $i$-band imaging. 
We adopt these corrections (0.65 for the quasar and 0.60 for the galaxy), but 
note that systematic effects may introduce additional (unaccounted for) 
uncertainty into these values.

Figure~\ref{fig:J0256LBTspec} displays the extracted spectra of both 
components, with slit loss corrections applied. The quasar spectrum has 
typical emission and absorption features for a high-redshift quasar. The 
galaxy spectrum is dominated by Ly$\alpha$ emission. Although it is not 
immediately apparent from Fig.~\ref{fig:J0256LBTspec}, we also detect 
continuum flux from the companion galaxy. Figure~\ref{fig:J0256LBTcont2D} 
presents a collapsed version of the 2D spectrum, binning pixels to enhance 
the $S/N$ until the weak continuum of the companion is apparent 
($S/N~\sim~0.5~{\rm pix}^{-1}$). The non-zero continuum level is also evident 
in the extracted 1D spectrum.

\subsection{Analysis}\label{sec:j0256anal}

\begin{figure}
 \epsscale{1.15}
 \plotone{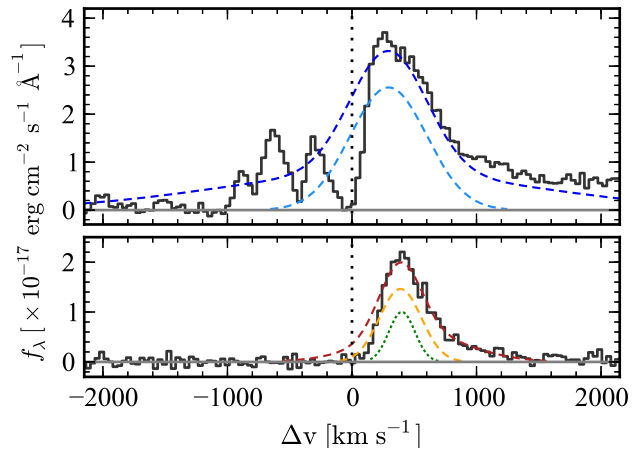}
 \caption{Extracted one dimensional LBT/MODS spectra showing the 
  Ly$\alpha$ emission regions of the quasar (upper panel) and companion galaxy
  (lower panel). The slit loss corrections have been applied, but these spectra 
  are not smoothed or binned.  Zero velocity is defined using the redshift
  obtained from the quasar's \ion{O}{1} emission line. 
  The dashed lines show the multiple Gaussian fits for both objects 
  (see Table~\ref{tab:J0256specfit}), with the lower line showing the narrow
  component fit and the upper line the sum of the narrow and broad components.
  The fit to the quasar line profile underestimates the total flux, especially 
  in the presence of the strong absorption features blueward of the line peak.
  The fit to the galaxy line profile captures the red wing that extends
  over $\sim1000~\kms$.
  The observed peaks of the quasar and galaxy Ly$\alpha$ lines are separated 
  by $\sim150~\kms$, although the intrinsic emission peak for the quasar may
  be significantly bluer and strongly absorbed, implying a larger velocity
  offset between the two objects.%
  The instrumental resolution is $\sim230~\kms$ (green dotted line in 
  lower panel).
 \label{fig:J0256LBTspecLyA}
 }
\end{figure}

Based on the MMT spectrum alone we considered the possibility that the 
companion represents a secondary image of a single source quasar due to 
gravitational lensing, and is visible only in Ly$\alpha$ emission as this 
was the strongest feature in the quasar spectrum. However, the quasar's 
\ion{C}{4} line is sufficiently strong that it should have been detected from 
the companion galaxy if its spectrum were simply scaled from the quasar's 
spectrum. This conclusion is even stronger based on the higher 
$S/N$ LBT observations: the flux ratios of the two spectra at the peak of 
the Ly$\alpha$ emission are $\sim$2:1, compared to a ratio at the 
\ion{C}{4} peak of $\ga$10:1 and a continuum ratio of $\sim$2.5:1.
Furthermore, the companion galaxy is resolved in the LBT imaging, and no 
candidate lens galaxy is apparent.  Finally, the detailed Ly$\alpha$ 
profiles show clear differences (see Figure~\ref{fig:J0256LBTspecLyA}).
Thus we reject the lensing hypothesis for this object.

Our main results are the detection of strong Ly$\alpha$ emission as well
as rest-frame UV continuum from the companion galaxy.
By fitting a power-law to the spectrum redward of Ly$\alpha$, we obtain
$f_{1500} = (4.7~\pm~0.2)~\times~10^{-19}~\fcgs$ (after applying the slit 
loss correction). This flux density corresponds to a continuum luminosity of 
$M_{1500} = -22.7$, or $\approx5.5~\Lstar$ at this redshift \citep{Bouwens+14}.
Although this value is high compared to local non-active
galaxies, objects with even higher UV luminosities have been found at $z\sim3$
\citep{Bian+12} and at $z\sim7$ \citep{Bowler+13}.

We measure a power-law slope of $\beta_{\lambda}=-1.1~\pm0.3$ 
($f_\lambda~\propto~\lambda^{\beta_\lambda}$) by fitting the UV continuum.
This value is quite uncertain given the low $S/N$ and the limited wavelength 
coverage. We fit the power law continuum of the quasar and obtain a slope 
$\alpha_\nu = -0.5$ ($f_\nu~\propto~\nu^{\alpha_\nu}$)\footnote{We follow 
the convention of measuring quasar power law slopes in frequency units and 
galaxy slopes in wavelength units.}. This agrees well with typical UV slopes 
measured for quasars \citep[e.g.,][]{VandenBerk+01} and suggests that the 
relative flux  calibration is reliable. A slope $\beta_\lambda = -1.1$ is 
fairly red, but consistent with measurements of galaxies at $z\sim5$, especially 
given that the spectral slope becomes redder at higher luminosities 
\citep{Bouwens+11}.

\begin{figure}
 \epsscale{1.15}
 \plotone{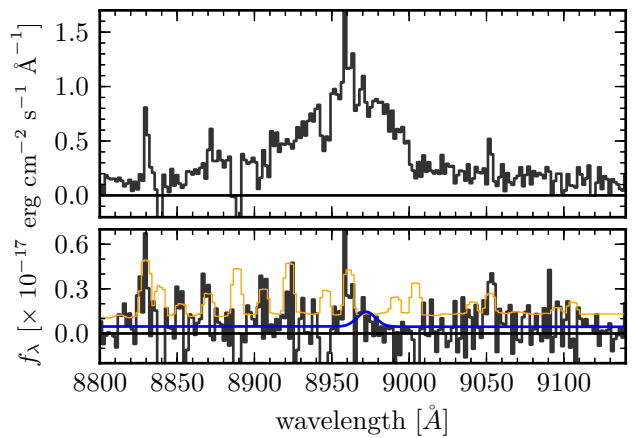}
 \caption{Extracted one dimensional LBT/MODS spectra showing the 
 \ion{C}{4} emission region (rebinned to 1.6~\AA~pix$^{-1}$ and corrected
 for slit losses). The upper
 panel displays the quasar spectrum and the lower panel the companion
 galaxy, along with the rms noise (thin orange line).
 The bottom panel also presents the fit to the galaxy continuum as a solid
 blue line, including a Gaussian profile to represent the $1\sigma$ upper 
 limit on the \ion{C}{4} emission (EW$_0~<~6$\AA). 
 This limit assumes the redshift obtained
 from the quasar's \ion{O}{1} line and a width of $200~\kms$ 
 (the quasar's \ion{C}{4} line is blueshifted, which is a common feature
 in quasar spectra [\citealt[cf.][]{Richards+11}]).
 \ion{C}{4} falls within a noisy portion of the OH forest and there are
 significant residuals from the sky line subtraction;
 however, it is evident the galaxy does not have strong \ion{C}{4} emission.
 The apparent feature at $\sim8960~\AA$ lies on a strong sky emission line;
 a similar feature is visible in the quasar spectrum and thus it is not likely
 to be real.
 \label{fig:J0256LBTspecCIV}
 }
\end{figure}

We measure a redshift of $z=4.789~\pm~0.005$ from a single Gaussian fit to 
the \ion{O}{1} emission line of the quasar. This is a low ionization 
transition which normally has smaller systematic redshift offsets than lines 
such as \ion{C}{4} \citep[e.g.,][]{Shen+11}; however, we adopt a conservatively
large uncertainty on the redshift to account for a possible shift of the 
\ion{O}{1} peak relative to the systemic velocity. We do not attempt to 
define a separate redshift for the companion galaxy, as the only detected
line is Ly$\alpha$, from which it is notoriously difficult to obtain a 
reliable redshift; however, the Ly$\alpha$ redshifts for the quasar and
galaxy agree to $<200~\kms$, suggesting that the true separation of the two
objects is not much larger than the projected separation of 12~kpc.

\begin{deluxetable}{lcc}
 \centering
 \tablecaption{Properties of \qsoa.}
 \tablewidth{0pt}
 \tablehead{
  \colhead{} &
  \colhead{QSO} &
  \colhead{Gal}
 }
 \startdata
RA (J2000) & 02:56:17.741 & 02:56:17.804\\
Dec (J2000) & +00:19:03.92 & +00:19:05.49 \\
$g$ (Stripe 82) & $>26.0$ & $>25.0$ \\
$r$ & $23.75 \pm 0.09$ & $>24.5$ \\
$i$ & $21.88 \pm 0.02$ & $23.6 \pm 0.2$ \\
$z$ & $21.93 \pm 0.10$ & $>22.5$ \\
$i$ (LBT) & $22.11 \pm 0.02$ & $23.62 \pm 0.09$ \\
$J$ (MMT) & $22.11 \pm 0.14$ & $>22.8 $ \\
$H$ & $21.43 \pm 0.10$ & $>23.2 $ \\
$m_{8700}$ & 24.7 & 25.6 \\
$M_{1500}$ & -23.6 & -22.7 \\
SFR(UV) [$\Msunyr$] & - & $13$ \\
$z_{{\rm Ly}\alpha}$\tablenotemark{a} & 4.794 & 4.796 \\
$f_{{\rm Ly}\alpha}$ [$\fcgs$] & - & $24 \times 10^{-17}$ \\
$L_{{\rm Ly}\alpha}$ [$\ergss$] & - & $58 \times 10^{42}$ \\
SFR(${\rm Ly}\alpha$) [$\Msunyr$] & - & $60$ \\
EW$_0$(Ly$\alpha$) [\AA] & 120 & 100 \\
\ldots broad & 53 & 38 \\
\ldots narrow & 65 & 42 \\
FWHM(Ly$\alpha$) [$\kms$] & 1380 & 690 \\
\ldots broad & 3010 & 970 \\
\ldots narrow & 720 & 400 \\
EW$_0$(\ion{N}{5}) [\AA] & - & $<$0.2 \\
EW$_0$(\ion{C}{4}) [\AA] & 75 & $<$5.7 
 \enddata
\label{tab:J0256specfit}
 \tablecomments{All photometry is on the AB system and corrected for Galactic
  extinction. Stripe 82 photometry for the quasar is derived from the coadded
  imaging. The $g$-band upper limit is the $3\sigma$ limit for point sources 
  in the field. We fit the companion galaxy with a 2D elliptical Gaussian 
  using \galfit\ and a PSF model derived from stars in the field using DAOPHOT.
  This model was applied to the other bands to derive $3\sigma$ upper limits.
  All flux-based quantities derived from the spectra have
  been corrected for slit losses (see~\S\ref{sec:lbtspec}).}
 \tablenotetext{a}{Redshift measured from the narrow component of the double Gaussian fit to the Ly$\alpha$ line.}
\end{deluxetable}

Figure~\ref{fig:J0256LBTspecLyA} shows the Ly$\alpha$ profiles of both
the quasar and companion galaxy. The companion galaxy has strong
Ly$\alpha$ emission, with $f_{{\rm Ly}\alpha} = 2.4~\times~10^{-16}~\fcgs$
and EW$_0~\sim100$~\AA. The equivalent width is not atypical when compared 
to high-redshift Lyman-alpha Emitters (LAEs) selected from narrow-band 
surveys \citep[e.g.,][]{Kashikawa+11}; however,
the line strength tends to decrease with continuum luminosity in these
objects; thus the galaxy stands out both for its bright continuum 
and the equivalent width of the Ly$\alpha$ emission line.
The line luminosity is $\sim6~\times~10^{43}~\ergss$; if the Ly$\alpha$ 
emission were entirely due to star formation, this would correspond to
SFR(Ly$\alpha$)$~\sim~60~\Msunyr$ using the \citet{Kennicutt98} relation 
for H$\alpha$ and assuming Case B recombination. 
This is likely to be an underestimate
due to IGM attenuation and dust extinction, both of which would
dampen the observed Ly$\alpha$ line flux.

On the other hand, the Ly$\alpha$ line profile is unusual when compared
to LAEs. From Figure~\ref{fig:J0256LBTspecLyA} it is evident that the red
wing of the line profile extends over a range $|\Delta{v}|\sim1000~\kms$. 
A fit to the line with a single Gaussian results in FWHM$~\sim~700~\kms$; 
a  fit with a double Gaussian has a broad wing with 
FWHM$~\sim~1000~\kms$. \citet{OTS04} observed a similar feature
in the spectrum of one of the Ly$\alpha$ 
companions to \br. They attributed the broad linewidth
to outflows of neutral hydrogen driven by star formation. Given the high
estimated SFR from Ly$\alpha$ (roughly four times that of \br\ Ly$\alpha$-1)
this scenario is plausible.

In Table~\ref{tab:J0256specfit} we provide values obtained from Gaussian
fitting of the emission lines in both spectra. For the quasar
Ly$\alpha$ line we restrict the fitting to the red wing of the line in order
to mask the strong absorption features in the blue wing. Because of the
difficulty in fitting the Ly$\alpha$ profile we do not provide formal 
uncertainties; similarly, the \ion{C}{4} line is in a region with substantial
night sky line residuals and is thus difficult to fit reliably.

The Ly$\alpha$ emission from the quasar is unusually narrow. We measure
a FWHM of $\sim750~\kms$ from a single Gaussian fit to the line. 
This is the narrowest line of the 35 $z\sim5$ quasars from our MMT 
observations from which we can measure the Ly$\alpha$ line (the median is 
$\sim2000~\kms$ with a standard deviation of 
$\sim500~\kms$; only one other object has FWHM$~<1000~\kms$).
The Ly$\alpha$ line alone would qualify this object as a Type II quasar 
candidate  according to the criteria of \citet{Alexandroff+13}. On the 
other hand, the \ion{C}{4} line is broader (FWHM~$\sim2000~\kms$) and the UV
continuum is strong ($M_{1450}~\approx~-24$), indicating at best a moderate
amount of nuclear extinction. It is evident from 
Figure~\ref{fig:J0256LBTspecLyA} that the Ly$\alpha$ line has multiple
strong absorption features; this likely explains the narrowness of the line. 
We re-fit the line using only the red wing and obtain FWHM$~\sim1400~\kms$. 
Although we do not attempt to model the Ly$\alpha$ absorption in detail,
the implication is that \qsoa\ has a high covering fraction of neutral gas 
along the line of sight when compared to quasars at a similar redshift. 
A virial estimate for the black hole mass from the \ion{C}{4} line width
using the relation of \citet{VP06} gives a mass of 
$\approx 1.2 \times 10^8~M_\sun$ and an Eddington ratio of 
$\lambda \approx 0.9$, indicating that the quasar is in a phase of rapid
growth.

No rest-frame UV emission lines other than Ly$\alpha$ are detected from the 
companion galaxy ; thus it is unlikely to be an AGN. Figure 
\ref{fig:J0256LBTspecCIV} shows the \ion{C}{4} emission region of both the
quasar and the companion. We obtain an upper limit on the
\ion{C}{4} emission from the companion galaxy by using the redshift obtained
from the quasar \ion{O}{1} line to set the wavelength and a fiducial linewidth 
of $\sigma_{\rm v} = 200~\kms$. This limit is EW$_0 < 6$~\AA\ ($1\sigma$),
whereas nearly all of the Type II quasar candidates in 
\citet{Alexandroff+13} have EW$_0$(\ion{C}{4}) $>$ 10~\AA.
Additionally, no \ion{N}{5} emission is detected from the galaxy, to a limit 
of EW$_0$(\ion{N}{5}) $<$ 0.2~\AA.
These properties are similar to the Ly$\alpha$-1 
companion galaxy of \br, which \citet{Williams+14} found to have ratios of 
\ion{C}{4} flux to Ly$\alpha$ flux $<0.033$ and \ion{N}{4}/Ly$\alpha$ $<0.019$,
compared to $>0.2$ and $\ga 0.1$, respectively, for typical 
Seyfert galaxies and obscured quasars. These ratios are $\approx0.05$ and
$<0.001$ for the \qsoa\ companion, suggesting that AGN photoionization plays 
at best a minor role.

\section{\qsob}\label{sec:j0050}

The $z\sim6$ quasar \qsob\ was discovered as part of the 
Canada-France High-z Quasar Survey \citep[CFHQS;][]{Willott+10}.
Compared to the bright $z\sim6$ quasars discovered in the SDSS main survey 
\citep[e.g.,][]{Fan+01PI}, the  CFHQS quasars are selected from smaller area, 
deeper imaging from various CFHT survey fields, and thus tend to be fainter. 
However, \qsob\ is one of the brightest quasars in the CFHQS, with 
$z_{\rm AB}=20.5$.  It is also one of the most distant objects, with 
$z=6.253 \pm 0.003$ measured from the \ion{Mg}{2} line \citep{Willott+10b}.
The \ion{Mg}{2} line was also used for a virial estimate of the BH mass,  
$M_{\rm bh} = 2.6~\times~10^{9}~M_\sun$. The mass and Eddington ratio 
($\lambda_{\rm Edd} = 0.6$) are typical of bright $z\sim6$ quasars.

\subsection{HST Observations}

We have been performing a systematic search for evidence of gravitational
lensing among $z\sim6$ quasars through an HST SNAP program
(\#12184, PI: X. Fan). This program includes nearly all of the known $z\sim6$
quasars and utilizes WFC3/IR F105W imaging. Full results of the SNAP
study will be presented in a forthcoming work (McGreer et al., in prep). 
During the SNAP program \qsob\ was observed with two 180s exposures. 
It was immediately apparent that the resulting image was inconsistent with
a single PSF detection, as there was excess flux detected at $\sim1$\arcsec\ 
from the quasar position. However, the shallow, single-band SNAP observations 
did not constrain the nature of the excess emission, which could be 
attributed to 1) a foreground interloper, 2) a high-redshift galaxy, or 
3) a secondary lensed image of the quasar.

\subsubsection{HST Cycle 19 Imaging}

\begin{figure}
 \epsscale{1.15}
 \plotone{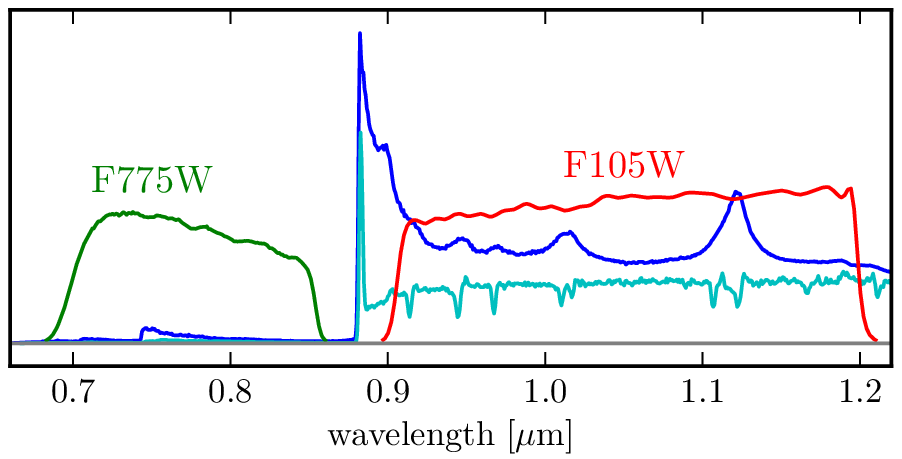}
 \caption{HST filter set for the \qsob\ observations. The labeled green and
 red lines show the F775W and F105W transmission curves, respectively.
 A template quasar spectrum \citep{VandenBerk+01} at $z=6.25$ is shown in 
 blue, and a template Lyman Break Galaxy \citep{Shapley+03} is shown in cyan. 
 Both templates have had Ly$\alpha$ forest absorption applied using
 a mean transmission spectrum from our simulations (\S\ref{sec:j0050field}).
 Ly$\alpha$ emission at this redshift falls between the two filters.
 \label{fig:J0050filtersed}
 }
\end{figure}

To further explore the nature  of the excess flux in the SNAP imaging we
obtained additional HST imaging of \qsob\ in Cycle 19 (\#12493, PI: 
I. McGreer). These observations consisted of two components: 1) three 
orbits of ACS/WFC imaging with the F775 bandpass, and 2) two orbits of 
WFC3/IR imaging with the F105W bandpass. This filter combination isolates 
objects at the redshift of the quasar by taking advantage of the strong 
spectral break feature introduced by the nearly saturated Ly$\alpha$ forest 
absorption at $z=6.25$. A single broadband color can provide strong evidence 
that a detected object is at high redshift if the filters straddle the Lyman
break. Figure~\ref{fig:J0050filtersed} 
displays the filter choices for the HST observations with SEDs of a 
high-redshift quasar and galaxy for reference. Neither of our 
filters includes Ly$\alpha$ emission at the redshift of the quasar.

The ACS/WFC imaging occurred on 2012 Jun 13. The three orbits were divided 
into two exposures, the first with four dither positions and the second with 
two dither positions. Images were processed using \astrodrizzle. 
First, the standard pipeline-reduced images were combined into two drizzled 
images, which were then aligned with the \textsc{tweakshifts} routine.
After propagating the calculated offsets to the input images they were
combined into a single drizzled image. The full integration time in the 
final mosaic is 8072s. We used pixfrac=0.8 and a scale of 0\farcs{03} for the 
output pixel grid \citep{Koekemoer+11}.

The WFC3/IR imaging occurred on 2012 Jun 8. A similar dither pattern was
employed. Although cosmic rays in individual images can be rejected by the 
continuous sampling of the detector array, we found some residual hot pixels 
in the individual images. We thus used \textsc{tweakreg} and 
\astrodrizzle\ in a single iteration with cosmic ray rejection enabled
to produce a final drizzled image combining all six individual images, with 
a total integration time of 5018s. We used pixfrac=0.8 and a scale of 
0\farcs{06} for the final pixel grid.

We found small ($<1$\arcsec) offsets between the final ACS and WFC3 mosaics 
when compared to SDSS imaging over the same area. We thus aligned the ACS 
image to the SDSS DR9 \citep{DR9} $i$-band image of the field using 
\textsc{tweakshifts}, and then aligned the WFC3 image to the corrected ACS
image. The final astrometric accuracy is $<~0.1$\arcsec\ when compared to 
SDSS.

\begin{figure*}
 \epsscale{1.15}
 \plotone{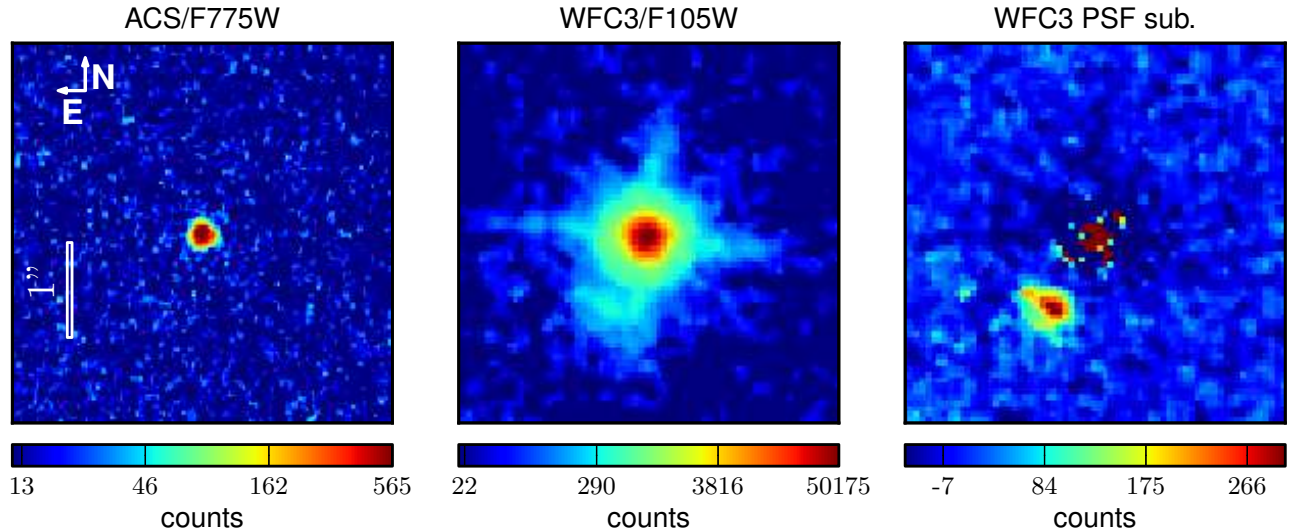}
 \caption{HST imaging of \qsob. The leftmost panel displays the ACS image,
 and the middle panel the WFC3 image. Both are shown with a log stretch
 and with the sky background calculated by \astrodrizzle\ removed.
 The rightmost panel shows the WFC3 image after subtraction of the PSF,
 and has been rescaled and placed on a linear stretch to highlight the galaxy.
 The companion has $Y_{105} = 25.0$ and is at a separation of 0.6\arcsec\ 
 from the quasar.
 \label{fig:J0050HST}
 }
\end{figure*}
\begin{figure}
 \epsscale{1.15}
 \plotone{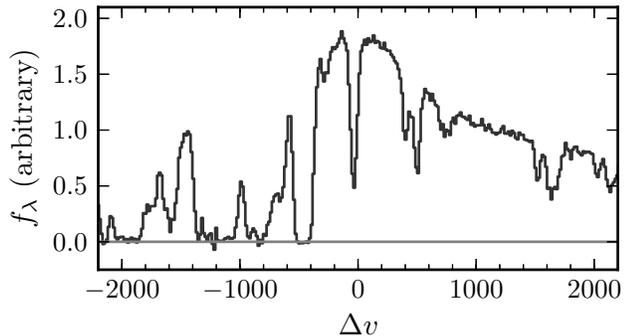}
 \caption{Portion of a 1.7~hr Keck ESI spectrum \citep{Becker+11} with 
 resolution $\sim50~\kms$ covering the Ly$\alpha$ emission region of \qsob. 
 Zero velocity is defined using the \ion{Mg}{2} redshift reported by 
 \citet{Willott+10b}. There is an absorption system close to zero velocity
 and a stronger system at $\sim-500~\kms$.
 \label{fig:J0050lya}
 }
\end{figure}

\subsection{Image Decomposition}\label{sec:j0050decomp}

Figure~\ref{fig:J0050HST} presents the HST imaging obtained for \qsob.
Within the central $\sim$1\arcsec\ the ACS image contains only a single
faint detection corresponding to the $z=6.25$ quasar, while the WFC3 
image includes both the bright quasar PSF component and an additional
extended component (a galaxy), consistent with the results from the 
shallow SNAP imaging.

We first construct PSF models in order to decompose the images. For both the
ACS and WFC3 images we generate empirical PSF models from stars within the
field using IRAF DAOPHOT tasks. Analysis of the ACS image is somewhat 
insensitive to the accuracy of the PSF model, as the quasar is relatively 
weak and the galaxy lies well beyond the extent of the quasar's 
light profile. On the other hand, the quasar is much brighter in the WFC3 
image, and positive flux from the wings of the PSF extends to the position 
of the galaxy. We selected eight isolated stars with high $S/N$ 
detections to construct the WFC3 PSF model. We also experimented with TinyTim 
PSF models, but found they were not an improvement over the empirical 
PSF \citep[cf.][]{Mechtley+12}.

We next use \galfit\ to model the quasar and galaxy in the HST 
images. The ACS image is fit with a single PSF component to represent the
quasar. The model for the WFC3 image includes both a PSF component for the
quasar and an exponential disk model for the galaxy. The exponential disk 
profile was selected empirically; we fix the axis ratio and position angle 
to values obtained from visually matching the galaxy profile in order to 
reduce the fit degeneracies. Finally, we find that the PSF subtraction is 
improved by masking the core of the quasar PSF during the fit. We mask the 
central 3$\times$3 pixels (0.18\arcsec). The results of the \galfit\ fitting 
are given in Table~\ref{tab:J0050galfit}.

The neighboring galaxy is not formally detected in the ACS image. We obtain
an upper limit for its flux by first subtracting the PSF component and then
fitting the residual image with a model for the galaxy.
The parameters for the galaxy model are fixed to
the values obtained from fitting the WFC3 image. We then increase the flux in 
the model galaxy until the fit results in $\Delta(\chi^2)=1$ compared to the
PSF-only fit. 
We adopt this value as the upper limit for the $i_{775}$ flux of the galaxy 
given in Table~\ref{tab:J0050galfit}. The limit obtained from this method, 
$i_{775}>26.9$, is bright compared to the average depth of the ACS
image. This results from a slight positive fluctuation at the  
position of the galaxy in the ACS image ($\sim+1\sigma$ per pixel for 
several neighboring pixels). However, this fluctuation appears to be 
associated with cosmic ray hits that appear in two out of the six ACS frames. 
We insert fake galaxies at random locations around the quasar position at the 
same radius of the neighboring galaxy and then repeat the process for deriving
an upper limit. We find that in general the limiting flux is $i_{775} > 28$.
Assuming that the positive flux is simply a noise fluctuation, the lower limit
on the color is $i_{775}-Y_{105}~\ga~3$ ($1\sigma$), consistent with a $z=6.25$ 
galaxy. However, we adopt the more conservative limit obtained at the 
galaxy's position in the image, yielding a color $i_{775}-Y_{105} > 1.8$.

\subsection{Analysis}\label{sec:j0050anal}

The original goal of the SNAP program was to search for gravitational
lenses, and one of the primary goals of the subsequent imaging was to test
the lensing hypothesis for \qsob. As one of most luminous quasars in the 
CFHQS, with $M_{1450} = -26.6$, and given the excess emission detected in the
SNAP imaging, \qsob\ was a prime candidate for a gravitationally lensed
quasar. However, the Cycle 19 HST imaging rules out the lensing hypothesis, 
at least at $\ga0.1$\arcsec\ scales, and thus suggests the measured luminosity 
is intrinsic. First, the secondary component is extended. While this 
could be a lens galaxy, no additional lensed quasar images are detected. The
limit obtained from the ACS imaging is even more stringent, with higher 
resolution and easier image decomposition. The quasar is detected at 
$\sim200\sigma$, and the detection limit for point sources is $i_{775}\sim29$, 
so flux ratios $\la60:1$ can be excluded at $\ga0.05$\arcsec\ separations.
These constraints effectively rule out all of the parameter space expected for
strong lensing of high-redshift quasars \citep{TOG84,CHS02,Richards+04};
even the inclusion of ellipticity or shear to the lens model will generally
result in strong lensing with multiple images that should be apparent with
HST's resolution and depth \citep{Keeton+05}.

\begin{deluxetable}{lcc}
 \centering
 \tablecaption{Properties of \qsob.}
 \tablewidth{2.5in}
 \tablehead{
  \colhead{} &
  \colhead{QSO} &
  \colhead{Gal}
 }
 \startdata
$\Delta{\alpha}$ & 0.00 & $+0\farcs45$ \\
$\Delta{\delta}$ & 0.00 & $-0\farcs74$ \\
$i_{775}$ & 24.032 $\pm$ 0.008 & $>$26.9 \\
$Y_{105}$ & 20.325 $\pm$ 0.003 & 25.06 $\pm$ 0.23 \\
$r_s$ & - & $0\farcs09 \pm 0\farcs03$ \\
$a/b$ & - & 0.4 \\
PA & - & $63.5^{\circ}$ \\
$M_{1500}$ & $-26.64$\tablenotemark{a} &  $-21.76$ \\
SFR(UV) & - &  27 $\Msunyr$ 
 \enddata
\label{tab:J0050galfit}
 \tablecomments{Measured quantities are from \galfit. The axis ratio ($a/b$)
 and position angle (PA) were fixed during fitting. PA is measured East of North.}
 \tablenotetext{a}{From \citet{Willott+10}.}
\end{deluxetable}

The neighboring galaxy is also unlikely to be a foreground interloper or even
a high-redshift galaxy unassociated with the quasar. The non-detection in the 
ACS imaging indicates an extremely red $i-Y$ color, akin to 
Lyman Break Galaxies (LBGs) at $z>6$. To examine the 
significance of the $i-Y$ color, we compare to the CANDELS GOODS-South 
compilation of \citet{Guo+13}, which includes ACS-$i_{775}$ imaging from 
GOODS \citep{Giavalisco+04} and WFC3-$Y_{105}$ imaging from CANDELS 
\citep{Grogin+11}.
We examine the distribution of $i_{775}-Y_{105}$ colors for objects within
the GOODS-S area with coverage in both filters, amounting to an area of
$\sim70~{\rm arcmin}^2$. We make conservative cuts of $Y_{105}<25$ and 
$i_{775}-Y_{105}>2.0$, finding a density of $\sim1~{\rm arcmin}^{-2}$. At this
density, the probability of finding a similar galaxy to the \qsob\ companion
within 1\arcsec\ of the quasar by chance is $\sim8~\times~10^{-4}$. This is 
strong evidence that the neighboring galaxy is indeed associated with the 
quasar. The projected separation between the two objects is 5.0 kpc if they
are at the same redshift.

The Ly$\alpha$ line falls outside of both the F775W and F105W bandpasses 
(Fig.~\ref{fig:J0050filtersed}). The detected emission is most likely 
dominated by UV continuum, while the presence of strong Ly$\alpha$ emission 
-- as in the case of \qsoa\ -- is unconstrained by our observations. The 
galaxy is resolved in the WFC3 image, with a fitted disk scale length of 
0\farcs{09}, or 0.5~kpc at $z=6.25$. The derived UV luminosity is 
$M_{1500}=-21.8$, which is $\approx 5\Lstar$ at this redshift 
\citep[e.g.,][]{McLure+13,Schenker+13}; and is comparable to the brightest 
galaxies found in ground-based surveys at this redshift 
\citep{Bowler+12}.

A possible explanation for the high UV luminosity of the \qsob\ companion 
galaxy is that it is powered by an AGN. At lower redshift, the fraction of 
galaxies hosting an AGN increases steeply with the host luminosity 
\citep[e.g.,][]{Juneau+13}. In fact, the derived UV luminosity of the \qsob\ 
companion is comparable to that of Seyfert galaxies. The possibility of an 
AGN is intriguing, as it would imply two massive black holes at $z=6.25$ 
separated by $<10$~kpc and potentially merging in a brief timescale. However, 
the currently available data do not provide any indicators to assess the 
presence of a low-luminosity AGN in the companion galaxy, other than to note 
that its morphology does not appear to be dominated by a central point source. 
Obtaining ground-based spectroscopy of the companion would be challenging, 
given its small separation from the nearby luminous quasar and its faintness.

\begin{figure}
 \epsscale{1.15}
 \plotone{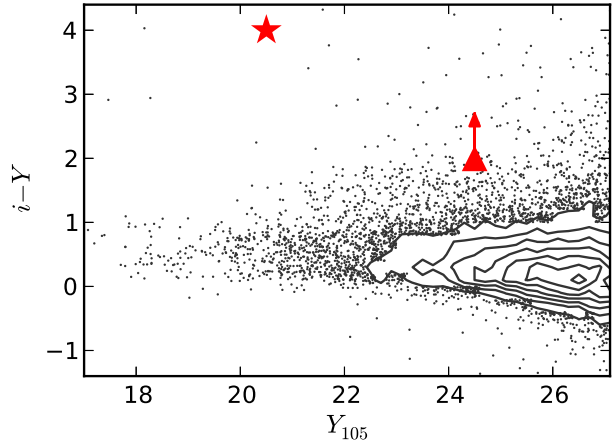}
 \caption{The distribution of $i-Y$ colors for objects from the GOODS-S
 catalogs of \citet{Guo+13}, shown as black contours and points. The area
 covered by the combined F775W/F105W imaging is $\approx14\times$ greater
 than for our single WFC3 pointing.
 The position of the quasar in this plot is represented by a red star,
 while the position of the companion galaxy (with the conservative lower 
 limit on the $i-Y$ color) is given by the triangle. The very red
 color of the quasar may be affected by the contribution of the \ion{C}{4} 
 line to the F105W bandpass.
 \label{fig:J0050imY}
 }
\end{figure}

\subsection{Search for additional neighboring galaxies}\label{sec:j0050field}

Our two-band imaging was designed to provide information on the excess
flux detected $<$1\arcsec\ from the quasar, but also allows
color selection of neighboring galaxies over the entire field. We thus search
for associated galaxies up to $\sim2$\arcmin\ from the quasar. The total area
of overlap between the ACS and WFC3 imaging is 4.9~arcmin$^{2}$. We construct
matched catalogs by first aligning the ACS image to the WFC3 image with 
\astrodrizzle\, and then executing \sextractor\ in dual-image mode with the 
WFC3 image for object detection.

We use simulations of the Lyman-$\alpha$ forest \citep[see][]{McGreer+13} 
and a simple power-law model for the UV continuum from galaxies with a 
range $-2.5 < \beta_\lambda < -1$ to estimate the range of galaxy colors 
expected at $z=6.25$\footnote{This is a much {\em narrower} selection than 
used in past surveys around $z\sim6$ quasars, which typically adopted a 
single, broad color cut that encompasses a large volume at $z>6$.}. From this 
approach the typical color is $i_{775} - Y_{105} \approx 3.1$, with 
$i_{775} - Y_{105} > 2.8$ for all simulated galaxies. The quasar has 
$i_{775} - Y_{105} = 3.7$. We estimate a $3\sigma$ detection limit in the 
ACS image of $i_{775}=28.2$ in a 0\farcs{24} aperture, though the same limit 
is somewhat weaker using the elliptical apertures defined for the WFC3 
detections (i.e., \sextractor\ MAG\_AUTO), for which we obtain 
$i_{775} \approx 27.8$. We thus survey for associated galaxies by searching 
for $i_{775}$-dropouts with $Y_{105}<25.0$; more specifically, we require 
$i_{775} - Y_{105} > 2.8$ to the detection limit of $i_{775} \approx 28$.

We identify four candidates from the \sextractor\ catalogs with these criteria 
(excluding the \qsob\ quasar and companion galaxy). Of those objects, three
are near the selection limit ($Y_{105} \approx 25$) and visual inspection
of the ACS images shows they are weakly detected in the optical band. We do
not consider them to be good candidates for $z \sim 6.25$ galaxies. The
final object has $Y_{105}=24.24 \pm 0.05$ and is also weakly detected in the
ACS image. Using a circular 1\arcsec\ diameter aperture instead of the 
MAG\_AUTO photometry, which roughly matches the size of the object in the 
WFC3 image, we obtain $i_{775} - Y_{105} \approx 1.9$. We also do not 
consider this object to be a good candidate for an associated galaxy.
From visual inspection we identify one object 11\farcs{5} from the quasar
that appears to be a true optical dropout; however, with $Y_{105}=25.4$ it 
is too faint for the color to be conclusive ($i_{775} - Y_{105} \ga 2.6$).

We also broaden our search to all optical dropouts to a depth of
$Y_{105}=27$. The ACS non-detections do not provide strong constraints on
the color for these objects. However, we can compare the number counts of
the faint WFC3 detections to those from the GOODS-S catalogs from CANDELS
described in \citet{Guo+13}. We restrict the GOODS-S area to regions covered
by both ACS/F775W and WFC3/F105W with integration times in each band greater 
than the integration times in our observations. We then consider any GOODS-S 
object with $i_{775}>28$ a ``dropout'', as it would be below the detection
limit of our ACS imaging. We find no excess of such objects in our imaging
after comparing to the number density obtained from the GOODS-S data.

The fact that we see no excess of associated galaxies on large scales around
the quasar \qsob\ is consistent with previous HST results that did not find
overdensities to be ubiquitous among high-redshift quasars 
\citep[e.g.,][]{Kim+09}. However, using F105W for detection restricts us to 
objects with bright rest-frame UV continuum; we are not sensitive to obscured 
sources or weak continuum sources with strong Ly$\alpha$ emission as the line 
falls outside our bandpass. On the other hand, a depth of $Y_{105}=27$ 
corresponds to $\approx L^*$ at  this redshift \citep{McLure+13,Schenker+13}; 
thus a strong overdensity of massive, star-forming galaxies should have been
detectable with our observations.

\section{Discussion}\label{sec:discuss}

\subsection{What powers the observed emission?}

The spectra of both quasars include strong Ly$\alpha$ absorption features near
the quasar redshift (Figures~\ref{fig:J0256LBTspecLyA}~and~\ref{fig:J0050lya}).
Such features are not uncommon in high-redshift quasar spectra, and may be 
indicative of a large reservoir of cold neutral gas surrounding the quasar. 
Further evidence for a high covering fraction of neutral gas in \qsoa\ comes 
from its relatively weak and narrow Ly$\alpha$ line. Such halos may give rise 
to fluorescent Ly$\alpha$ emission \citep{HR01} powered by the quasar's 
ionizing continuum. Could the companions be dense clouds reprocessing the 
ionizing flux from the quasars, rather than self-luminous systems?

We consider the contribution of fluorescent Ly$\alpha$ emission to the observed 
spectrum of the \qsoa\ companion galaxy by determining whether the incident
ionizing flux from the quasar onto the galaxy is sufficient to power its strong
Ly$\alpha$ flux. We adopt the projected separation between the two objects and 
a radius of 3~kpc for the companion (roughly its measured extent from the LBT 
image). We further assume that the quasar UV emission is isotropic and has a 
typical UV power law continuum \citep{SSD12}. The incident ionizing photon flux on the galaxy is then $\sim4\times10^{54}~\ergss$. If every photon resulted in 
a Ly$\alpha$ photon, the resulting Ly$\alpha$ luminosity would be 
$\sim6~\times~10^{43}~\ergss$, which is roughly equal to the measured 
luminosity. In other words, the observed strong Ly$\alpha$ emission could be 
due entirely to fluorescence if {\em all} of the quasar's ionizing photons 
incident on the galaxy are re-emitted as Ly$\alpha$ photons by the galaxy, 
whereas only $\sim$60\% of ionizing photons should result in Ly$\alpha$ 
photons even for a single optically thick cloud \citep{GW96}.

The \qsoa\ companion galaxy also has strong continuum emission, with 
$\nu{L}_\nu \approx 10^{45}~\ergss$ at rest-frame 1500~\AA, so its spectrum 
cannot be purely fluorescent.  Similarly, for \qsob\ the observed emission 
arises only from continuum redward of Ly$\alpha$. The lack of other emission 
lines in the spectrum of the \qsoa\ companion further demonstrates that neutral 
gas within its ISM is not being ionized by the nearby quasar [e.g., 
\citealt{Gunn71,Filippenko85}; see also the case of Hanny's Voorwerp, a cloud 
of gas illuminated by a (now dormant) quasar that has strong high ionization 
lines \citep{Keel+12}].
However, it is intriguing to consider that, at least in the case of
\qsoa, {\em some} of the strong Ly$\alpha$ emission may arise from fluorescence 
in a neutral halo of metal-poor gas surrounding the galaxy. This 
interpretation  is consistent with the larger SFR inferred from the 
Ly$\alpha$ emission than the UV continuum (about a factor of four, see 
Table~\ref{tab:J0256specfit}).

The companion galaxies may themselves be AGN. Indeed, in merger-driven models 
for high-redshift quasar triggering \citep[e.g.,][]{Li+07} it is expected that 
the progenitor galaxies each carry their own massive black hole, although it
may not be likely for both to be active at the same time. We argued in 
\S\ref{sec:j0256anal} that the lack of AGN lines in \qsoa\ argues against an 
AGN in this object. The constraints on AGN activity in the \qsob\ companion 
are weaker, although we noted in \S\ref{sec:j0050anal} that its observed 
emission is not nuclear-dominated.

While we cannot conclusively rule out AGN activity in either system, the
current observations are consistent with both galaxies being powered by
internal star formation. We next discuss the implications of finding two 
unusually bright galaxies in the vicinity of high-redshift quasars.

\subsection{Are these major mergers?}

The presence of bright galaxies within $\sim$10~kpc of two luminous, 
high-redshift quasars is highly suggestive of ongoing major mergers
in both systems. Far-IR measurements of \ion{C}{2} line widths of $z\sim6$ 
quasars have shown that their dynamical masses are roughly 
$\sim10^{10-11}~M_{\sun}$ \citep{Wang+13}. \ion{C}{2} dynamical masses
may underestimate the full mass of the stellar bulge if the 
line emission is mainly concentrated in the center of the host galaxy; 
however, CO observations further indicate masses $>10^{10}~M_{\sun}$ in 
molecular gas alone \citep{Wang+10}. This result is consistent with an 
expectation that the hosts of high-redshift quasars are massive, with the 
archetypal case being J1148+5251 at $z=6.4$ \citep{Walter+03}.

Next, we consider the galaxy companions. According to the $z=5.7$ Ly$\alpha$ 
luminosity function from \citet{Kashikawa+11}, the \qsoa\ companion galaxy is 
at $\sim6\Lstar$, and thus on the steeply falling bright end of the luminosity 
function.
It can be compared to {\it Himiko}, an extremely bright LAE discovered in the 
Subaru Deep Field \citep{Ouchi+09}. The \qsoa\ companion galaxy has many 
features in common with {\it Himiko}: it is resolved in ground-based imaging, 
and has a comparable Ly$\alpha$ luminosity (if not somewhat  greater), 
equivalent width, and UV continuum luminosity. Using multiwavelength SED 
fitting,  \citet{Ouchi+09} inferred a stellar mass of 
$\sim4\times10^{10}~M_\sun$ for {\it Himiko}.

We can compare \qsob\ to field LBGs selected at a similar redshift.
\citet{CurtisLake+13} examined luminous ($L>1.2L^*$) galaxies with 
spectroscopic redshifts $5.5<z<6.5$ and found that the stellar masses lie
in the range $10^{9}~<M_*~<10^{10}~M_{\sun}$, with typical stellar population
ages of $50\mbox{--}200$~Myr. At $\approx5~L^*$, the companion to \qsob\ is
comparable to the most extreme objects in the \citet{CurtisLake+13} sample,
and thus likely lies at the higher mass end of this distribution.
Similarly, the stellar mass-UV luminosity relation of \citet{Stark+09} derived
at $z\sim4$ would predict a mass of $M_*~\sim~10^{10}~M_{\sun}$ at this
luminosity; this relation shows little or no evolution to higher redshift
\citep[e.g.,][]{CurtisLake+13}. 

Thus, comparison to results from high-redshift surveys implies that 
both the quasar host galaxies and the companion galaxies have masses
$>10^{10}~M_\sun$. 
The small projected separations between the quasars and companion galaxies
($\la 10$~kpc) in both cases place the quasar host and companion within the 
virial radius of a $10^{11}\mbox{--}10^{12}~M_\sun$ halo (the mass range of
a halo likely to host a luminous quasar at this redshift), such that 
their eventual coalescence seems inevitable.
A likely conclusion, then,  is that in both cases we are witnessing 
extreme major merger events, potentially fueling luminous quasar and powerful 
starburst activity simultaneously.

\subsection{How unusual are these systems?}

Both of the systems discussed here were discovered serendipitously and
it is not straightforward to draw conclusions about how common or rare
they may be. However, we can ask whether similar systems would have been
detected in the course of our surveys.

\qsoa\ was discovered in slit spectroscopic observations. We observed
55 quasars at $z\sim5$ with the 180\arcsec\ longslit on MMT Red Channel during
the course of our Stripe 82 survey \citep{McGreer+13}.
Roughly half were observed with a 1\arcsec\ slit, the other half with
a 1.5\arcsec\ slit, so we use an average slit width of 1.25\arcsec. 
If we consider neighbors within 20~kpc ($\sim$3\arcsec), 
the fraction of area within that radius of the quasar subtended by the slit
is $\sim13$\% (ignoring slit losses for miscentered objects).
Thus we would have detected 1 in $\sim7.5$ galaxy neighbors in our quasar
survey. Experimenting with a single exposure of \qsoa, if the Ly$\alpha$ 
emission were a factor of $\sim2$ fainter, it would have still been noticeable 
during the data reduction. From this crude analysis we conclude that less than 
1 in $\sim55/7.5 \approx 7$ quasars in our $z=5$ survey have a companion with a 
Ly$\alpha$ flux within a factor of two of the \qsoa\ companion.

It is somewhat easier to draw statistical conclusions from the \qsob\ 
observations. A total of 29 $z\sim6$ quasars were observed during the HST
SNAP survey. Brighter quasars ($z_{\rm AB} < 20.5$) had a total exposure time
of 300s, while the fainter objects were exposed for 1200s. \qsob\ was in the 
bright sample and its companion galaxy was detected near the limit of the 
imaging; the faint sample could reach galaxies $\sim1.6$ times fainter. Thus
we could have detected companion galaxies to a limit of $\approx5\Lstar$ for
the 23 bright objects observed, and $\approx2\Lstar$ for the 6 faint objects.
One other bright quasar has excess emission in the SNAP imaging
and is scheduled for further observations in Cycle 21; as part of the search
for gravitational lenses we can rule out any other companions at separations
$<3$\arcsec. Thus the incidence of UV-bright galaxies associated with quasars 
at $z\sim6$ is $\la 2/29$ for $\ga5\Lstar$ galaxies and $<1/6$ for 
$2~\la~L~\la~5\Lstar$ galaxies.

At least within the context of our observations, bright companions to
high redshift quasars are uncommon. We conclude with some final thoughts
related to the incidence of such systems.

\section{Conclusions}\label{sec:conclude}

We have identified two galaxies in close proximity to high-redshift quasars.
The first object is located 1\farcs{8} from a $z=4.9$ quasar (\qsoafull) and 
was discovered serendipitously based on excess Ly$\alpha$ emission present in 
the spectroscopic slit during the course of an MMT survey of $z\sim5$ quasars.
Imaging and moderate depth optical spectroscopy with the LBT confirms that 
the galaxy is bright ($i=23.6$) and has strong Ly$\alpha$ emission at the 
quasar redshift as well as UV continuum emission. The second companion galaxy 
is $<1$\arcsec\ from a highly luminous $z=6.25$ quasar (\qsobfull). HST 
imaging demonstrates that it is a relatively bright ($Y=25$) optical dropout 
and highly likely to be at the same redshift as the quasar.

Both galaxies are among the most luminous galaxies known at high redshift,
with $M_{1500}=-22.7$ for the \qsoa\ companion and $M_{1500}=-21.8$ for the
\qsob\ companion. We have considered possible sources for their emission, 
including internal AGN or fluorescence from the nearby quasars, and for both 
objects conclude that the observed emission is likely dominated by star 
formation activity within the galaxies. The coincidence of highly
star-forming galaxies near luminous quasars is broadly consistent with
the scenario where high-redshift quasars are fueled by major mergers.

This study is not the first to find close companion galaxies to high-redshift 
quasars. The first galaxy to have a reported spectroscopic redshift $z>2$ was 
a companion to the $z=3.2$ quasar QSO PKS 1614+051 with a separation of 
7\arcsec\ \citep{Djorgovski+85}. This galaxy was selected from narrow-band 
Ly$\alpha$ imaging of the quasar, and was only instance of a companion out of 
five objects surveyed. The $z=4.7$ quasar \br\ was found to have multiple 
companions within $\sim5$\arcsec, two discovered via Ly$\alpha$ 
\citep{HME96,Petitjean+96} and one in the sub-millimeter \citep{Omont+96}.
While searching for CO(2-1) line emission from the host galaxies of five 
quasars at $z\sim6$, \citet{Wang+11} found one clear detection of a companion 
galaxy 1\farcs{2}  from a $z=6.18$ quasar, and a marginal $\sim2\sigma$ 
detection of extended CO emission $\sim$0\farcs{8} from a $z=5.85$ quasar. 
These studies either targeted quasar fields under the expectation they would
point to a local galaxy overdensity, or targeted quasar hosts while being 
sensitive to any nearby companions.

One can then ask why companion galaxies are not observed more frequently,
given the observational attention paid to high-redshift quasars and the 
hypothesis that major mergers play a key role in triggering high-redshift 
quasars. One explanation is that the merging timescales are brief, so that
observing quasar-galaxy pairs would be rare. Indeed, the ``blowout'' phase 
that leads to an unobscured quasar may occur only after the progenitor
galaxies have fully coalesced \citep[e.g.,][]{Hopkins+06}. In the scenario
where successive galactic encounters fuel the growth of high-redshift 
quasars, the duty cycles of unobscured activity for both star formation (in
the progenitor galaxies) and black hole growth may be sufficiently short 
(relative to the quasar lifetime) that the  probability of observing the two 
simultaneously at rest-frame UV/optical wavelengths is low. Larger, more 
sensitive surveys at FIR  wavelengths with ALMA would better address this 
question. At rest-UV wavelengths, \citet{Jiang+13} found that 
$\sim50$\% of bright $z\sim6$ galaxies show evidence for recent mergers or 
interactions in HST observations. However, observing such faint features in
luminous quasar host galaxies would be exceedingly difficult.

A final speculation motivated by our observations is that the proximity of a 
UV-bright galaxy to a rapidly growing supermassive black hole at high redshift 
may be indicative of a form of {\em positive} feedback. \citet{Dijkstra+08}
outline a model where massive seed black holes for $z\sim6$ quasars form
at even higher redshifts ($z\sim10$) from close pairs ($\la 10$kpc) of dark 
matter halos. In this model, one halo hosts a bright star-forming galaxy with 
a strong Lyman-Werner band flux that photo-dissociates $H_2$ molecules in the 
neighboring halo, preventing it from cooling below the atomic cooling 
threshold. The result is direct collapse of the neighboring halo gas into a 
massive ($\sim10^{4\mbox{-}6}~M_{\sun}$) black hole, providing a seed 
mechanism for supermassive black holes. However, the original ionizing source 
at $z\sim10$ would likely coalesce with the quasar host by $z\sim6$, and not 
explain our observations in which a bright companion galaxy is contemporaneous 
with an already $\ga10^8~M_\sun$ black hole. On the other hand, observing
a companion galaxy to a  quasar well after the initial formation of the black 
hole is consistent with the picture where multiple
mergers are needed to grow high-redshift quasars \citep[e.g.,][]{Li+07},
drawing at least some connection between the model of \citet{Dijkstra+08} and
our observations.

\textbf{Acknowledgements}

The authors would like to thank George Becker for providing the ESI spectrum
of \qsob. IDM would like to thank Desika Narayanan and Dan Stark for helpful
discussions.

Support for programs \#12184 and \#12493 was provided by NASA through a 
grant from the Space Telescope Science Institute, which is operated by the 
Association of Universities for Research in Astronomy, Inc., under NASA 
contract NAS 5-26555. IDM and XF acknowledge additional support from
NSF grants AST 08-06861 and AST 11-07682.

The LBT is an international collaboration among institutions in the United States, Italy, and Germany. LBT Corporation partners are: The University of Arizona on behalf of the Arizona University System; Istituto Nazionale di Astrofisica, Italy; LBT Beteiligungsgesellschaft, Germany, representing the Max-Planck Society, the Astrophysical Institute Potsdam, and Heidelberg University; The Ohio State University, and The Research Corporation, on behalf of The University of Notre Dame, University of Minnesota, and University of Virginia.
This paper used data obtained with the MODS spectrographs built with
funding from NSF grant AST-9987045 and the NSF Telescope System
Instrumentation Program (TSIP), with additional funds from the Ohio
Board of Regents and the Ohio State University Office of Research.

The MMT Observatory is a joint facility of the University of Arizona and the Smithsonian Institution.

{\it Facilities:}
 \facility{MMT (Red Channel spectrograph, SWIRC)},
 \facility{LBT (MODS)},
 \facility{HST (ACS,WFC3)},
 \facility{SDSS}
 
 \bibliographystyle{hapj}
\bibliography{hizqsocompanions}

\clearpage

\end{document}